\documentclass{iau}
\usepackage{graphicx}
\usepackage{natbib}
\bibpunct{(}{)}{;}{a}{}{,}

\title[The origin of magnetic fields in hot stars]
{The origin of magnetic fields in hot stars}

\author[Neiner et al.]   
{Coralie Neiner$^1$, St\'ephane Mathis$^{2,1}$, Evelyne Alecian$^{3,1}$, Constance Emeriau$^2$, Jason Grunhut$^4$, and the BinaMIcS and MiMeS collaborations}

\affiliation{$^1$LESIA, Observatoire de Paris, CNRS UMR 8109, UPMC, Universit\'e Paris Diderot, 5 place Jules Janssen, 92190 Meudon, France\\
email: {\tt coralie.neiner@obspm.fr}\\[\affilskip]
$^2$Laboratoire AIM Paris-Saclay, CEA/DSM - CNRS - Universit\'e Paris Diderot, IRFU/SAp Centre de Saclay, 91191 Gif-sur-Yvette, France\\[\affilskip]
$^3$UJF-Grenoble 1/CNRS-INSU, Institut de Plan\'etologie et d'Astrophysique de Grenoble (IPAG) UMR 5274, 38041 Grenoble, France\\[\affilskip]
$^4$ESO, Karl-Schwarzschild-Strasse 2, 85748 Garching, Germany
}

\pubyear{2015}
\volume{305}  
\pagerange{119--126}
\jname{Polarimetry: from the sun to stars and stellar environments}
\editors{K.N. Nagendra, S. Bagnulo, R. Centeno \& M. Mart\'inez Gonz\'alez, eds.}
\begin{document}

\maketitle

\begin{abstract} Observations of stable mainly dipolar magnetic fields at the
surface of $\sim$7\% of single hot stars indicate that these fields are of
fossil origin, i.e. they descend from the seed field in the molecular clouds
from which the stars were formed. Recent results confort this theory. First,
theoretical work and numerical simulations confirm that the properties of the 
observed fields correspond to those expected from fossil fields. They also
showed that rapid rotation does not modify the surfacic dipolar magnetic
configurations, but hinders the stability of fossil fields. This explains the
lack of correlation between the magnetic field properties and stellar properties
in massive stars. It may also explain the lack of detections of magnetic fields
in Be stars, which rotate close to their break-up velocity. In addition,
observations by the BinaMIcS collaboration of hot stars in binary systems show
that the fraction of those hosting detectable magnetic fields is much smaller
than for single hot stars. This could be related to results obtained in
simulations of massive star formation, which show that the stronger the magnetic
field in the original molecular cloud, the more difficult it is to fragment
massive cores to form several stars. Therefore, more and more arguments support
the fossil field theory. \keywords{stars: magnetic fields, stars: rotation,
stars: formation, stars: early-type} \end{abstract}

\firstsection 
\section{Introduction}

The MiMeS (Magnetism in Massive Stars) project showed that about 7\% of single
OB stars are magnetic \citep{neiner2011,wade2014}. In addition, a similar
proportion of A stars are known to be magnetic \citep{wolff1968,power2007}. The
magnetic fields of OBA stars have simple configurations, stable mainly oblique
dipoles, and their strengths range from $\sim$100 to $\sim$30000 G. More details
on the properties of magnetic fields in hot stars are presented in  Grunhut et
al. (these proceedings). Therefore, there seems to be a common origin of
magnetic fields in all hot (OBA) stars. This origin, however, has remained
unknown for a long time.

In cool stars, including the Sun, magnetic fields are generated and sustained by
a dynamo in the convective envelope \citep[e.g.][]{charbonneau2010,brun2004}. As
a consequence, the magnetic fields of cool stars are highly dynamic, and exhibit
variability on a very wide range of timescales. The internal structure of the
star, its rotation, and its accretion state can strongly influence the dynamo,
and ultimately set the broad properties of the magnetic field. The
dynamo-generated magnetic field, in turn, drives the mass loss and angular
momentum loss through magnetised winds and coronal mass ejection processes
\citep[e.g.][]{matt2012,reville2015}. Therefore, a complex interplay exists
between magnetic fields and rotation of cool stars during their whole evolution.

These properties are not observed in hot stars. They do not have a thick outer
convective envelope and a Sun-like dynamo can thus not develop. The origin of
their magnetic field must be found elsewhere.

\section{Dynamo fields?}

\subsection{Dynamo field in the convective core}

A hot star consists of a convective core, a radiative envelope, and a very thin
convective layer just below the surface. Like in the external convective
envelope of low-mass and solar-type stars, a dynamo takes place in the
convective core of intermediate-mass and massive stars. It generates and
sustains a magnetic field, because of the combined action of differential
rotation and turbulent helical flows  \citep[i.e. an $\alpha-\Omega$ dynamo
action; e.g.][]{brun2005}. However, the time needed for this field to reach the
surface and become visible is longer than the lifetime of the star
\citep{charbonneau2001, MGC2003}. Moreover, an $\alpha-\Omega$ dynamo would lead
to a correlation between the magnetic field properties and stellar rotation,
which is not observed. Therefore, even if such a core dynamo field exists, it is
not the one that we observe at the surface of hot stars.

\subsection{Dynamo field in the radiative envelope}

Over the last decade, various groups investigated the possibility of creating a
dynamo in the radiative envelope of hot stars
\citep[e.g.][]{spruit2002,zahn2007,arlt2011,rudiger2012,Jouve2014}. Like in
convective regions, the $\Omega$ effect, i.e. differential rotation, transforms
an initial axisymmetric poloidal field into an axisymmetric toroidal field.
Tayler's and other MHD instabilities, that can develop in radiation regions,
then transform this field into a field with a non-axisymmetric component
\citep[][]{Tayler1973,Markey1973,Brun2007}. To maintain the magnetic field, it
is then necessary to close the dynamo loop, by regenerating an axisymmetric
field from the non-axisymmetric field. \cite{spruit2002} proposed to regenerate
the axisymmetric toroidal field, while \cite{braithwaite2006} proposed to
regenerate the poloidal field. For this, both used the shear, but
\cite{zahn2007} showed that axisymmetric fields cannot be regenerated by the
shear alone. Instead, \cite{zahn2007} proposed to close the loop thanks to the
electromotrice force of the instability \citep[see also][]{rudiger2012}. While
this seems to work theoretically, numerical simulations have shown that this
dynamo is not excited or maintained.

Moreover, if an $\alpha-\Omega$ dynamo existed in the radiative envelope, a
correlation would exist between the rotation and the magnetic field properties.
Such a correlation is not observed in OB stars \citep{wade2014}. Consequently,
the possible production of a dynamo field in the radiative envelope of hot stars
must be rejected.

\subsection{Dynamo field in the sub-surface convection layer}

Hot stars have a very thin convective layer just below their surface.
\cite{cantiello2011} showed that a dynamo may develop in this layer. However,
the fields produced this way are of the order of 5 to 50 G for B stars, which is
much weaker than the magnetic fields observed at the surface of these stars.
Moreover, a magnetic field produced by sub-surface convection would likely have
a small-scale and time-dependent structure, while the observed fields are mostly
dipolar and stable. As a result, even if such a sub-surface dynamo field may
exist, it does not correspond to the ones observed at the surface of hot stars.

\section{Fossil fields}

\subsection{Fossil origin of magnetism in hot stars}

\begin{figure}[t]
\begin{center}
\includegraphics[width=\textwidth]{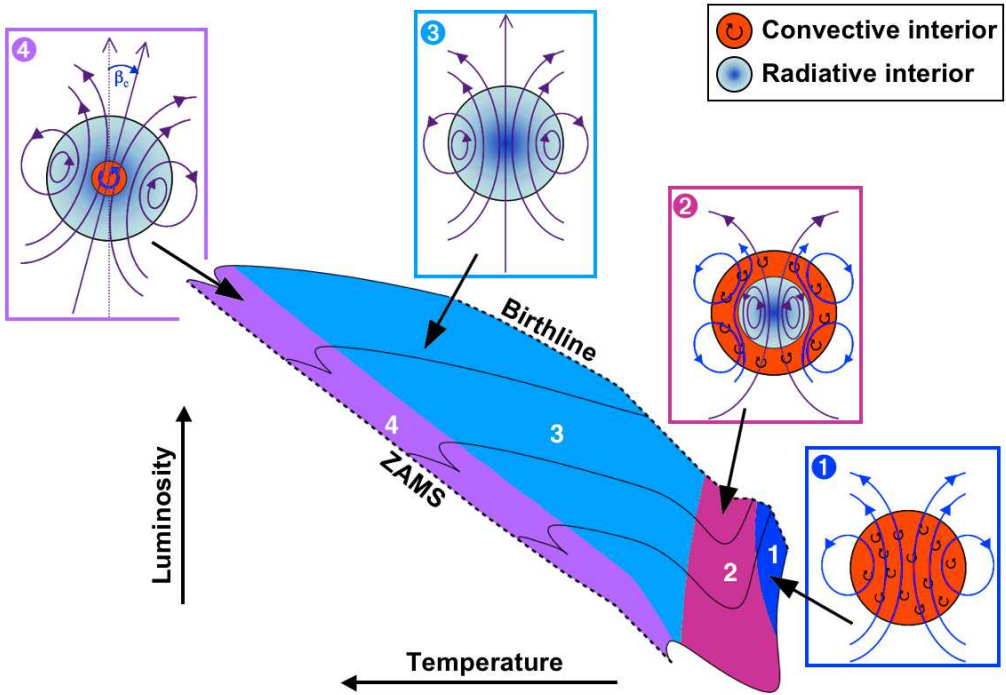} 
\caption{Diagram showing the pre-main sequence (PMS) from the birthline to the
ZAMS, with various evolutionary tracks from 1.2 to 8 M$_\odot$ shown with black
solid lines. The PMS is divided in 4 parts indicated with colours, showing 4
stages of the evolution of the structure of the stars and of their fossil magnetic field.}
\label{fossildiagram}
\end{center}
\end{figure}

During the formation of a hot star, the magnetic field present in the molecular
cloud can get trapped in the star as the cloud collapses. Fossil magnetic fields
are descendants from this seed field \citep{Mestel1999}. During the early stage
of the life of the star, when it is fully convective, this seed field can get
enhanced and sustained by a dynamo. As the radiative core forms and the
convective turbulence disappears in the center of the star, this dynamo field
relaxes onto a large-scale mixed (poloidal+toroidal) stable (possibly oblique)
dipole. Such relaxation processes have been observed in numerical simulations
\citep{braithwaite2004,braithwaitenordlund2006}. Moreover, theoretical work 
demonstrated that this mechanism results from a selective decay of the energy of
the system and of ideal MHD invariants such as magnetic helicity
\citep{duez2010}. Finally, to stay stable on long timescale, the field must have
a given ratio of the relative energies contained in its toroidal and poloidal
components \citep{braithwaite2009,duezbm2010}. The external convection zone then
disappears and the star becomes fully radiative with the dipolar fossil field
emerging at its surface. It is possible that the appearance of the convective
core, just before reaching the ZAMS, produces a (extra) tilt of the dipole and
explains why oblique dipoles are observed in basically all hot stars. Indeed,
from ASH 3D MHD simulations, \cite{featherstone2009} showed that the interaction
of a core dynamo with a fossil field in the envelope produces several effects:
it strengthens the core field, it makes the rotation of the envelope more rigid,
and it changes the orientation of the fossil field in the envelope. 
Figure~\ref{fossildiagram} shows a diagram of the evolution of fossil fields. 

In addition, \cite{alecian2013} showed that Herbig Ae/Be stars, which are the
precursors of the magnetic Ap/Bp stars, host magnetic fields with a similar
occurence rate and configuration to main sequence hot stars. This indicates that
the fields observed in hot stars are already present at the PMS phase.

Therefore, it is now well established that the magnetic fields of hot stars are
of fossil origin. However, the exact details of the creation and evolution of
these fields, from the molecular cloud to the main sequence, require further
investigations, in particular on the influence of rotation and stellar formation
conditions. 

\subsection{Impact of rotation on a fossil field}

Recent theoretical calculations showed that rotation modifies the internal
distribution of the magnetic flux and the stability of fossil magnetic
fields, while it does not modify their surface geometric configuration. Indeed,
fossil fields relax onto mixed dipolar configurations, no matter how fast the
star rotates \citep{emeriau2014}. However, as demonstrated by
\cite{braithwaite2013}, the time needed to reach equilibrium increases with
rotation. Therefore, it is probably more difficult for a rapidly rotating star
to reach a stable dipolar configuration. 

This would explain, in particular, why magnetic fields have not been directly
detected in classical Be stars, which rotate close to their breakup velocity
\citep[see][]{wade2014be}. However, rapidly rotating magnetic hot stars do
exist. This is the case, for example, of HR\,7355 \citep{oksala2010} and
HR\,5907 \citep{grunhut2012}. More investigations on the impact of rapid
rotation are thus needed.

When equilibrium is not reached, the star can still host a magnetic field, but
this field will most likely be very weak and on small scales
\citep{auriere2007,lignieres2014,braithwaite2013}. This kind of ultra-weak field
has been observed in Vega \citep{lignieres2009,petit2010} and in a few Am stars
\citep[e.g.][see also Blazere et al., these proceedings]{petit2011}.

\section{Lack of magnetic fields in hot binaries}

BinaMIcS \citep[Binarity and Magnetic Interactions in various classes of
Stars,][]{neiner2013,alecian2014} is an ongoing project that exploits binarity
to yield new constraints on the physical processes at work in hot and cool
magnetic stars. It rests on two large programs of observations with the ESPaDOnS
spectropolarimeter at CFHT in Hawaii and its twin Narval at TBL in France.
BinaMIcS aims at studying the role of magnetism during stellar formation,
magnetospheric star-star (and star-planet) interactions, the impact of tidal
flows on fossil and dynamo fields, its impact on mass and angular momentum
transfer, etc. 

In the frame of BinaMIcS, a large survey of magnetism in hot spectroscopic
binary systems with 2 spectra (SB2) has been undertaken. Out of $\sim$200 
observed SB2, including at least one star (and most of the time two stars) with
spectral type O, B or A in each system, none were found to host a magnetic
field, while the detection threshold was similar to the one used in the MiMeS
project on single hot stars. This lack of detections in $\sim$400 stars with
BinaMIcS, compared to the $\sim$7\% detection rate in $\sim$500 single stars
with MiMeS, is thus statistically significant: magnetism is less present in hot
binaries than in single hot stars.

This lack of magnetic stars in hot binaries might be related to results obtained
from simulations of star formation by \cite{commercon2011}. They found that the
more magnetic the medium is, the less fragmentation of dense cores there is. In
other words, when the medium is magnetic, it is more difficult to form binaries.
As a consequence, forming a binary with a fossil field is unlikely. 

Nevertheless, 6 SB2 systems hosting a magnetic OBA star are known to exist.
These are HD\,5550, HD\,37017, HD\,37061 (NU Ori), HD\,47129 (Plaskett's star),
HD\,98088, and HD\,136504 ($\epsilon$\,Lup). It is possible that magnetic hot
binaries still form sometimes. However, for these 6 systems, only one of the two
components is known to be magnetic, which is puzzling if the stars were formed
simultaneously. A possible explanation is that these binaries were formed in a
later stage of stellar evolution, e.g. by capture, from a magnetic single hot
star and non-magnetic hot star. 

\section{Conclusions}

The magnetic fields of single hot stars is of fossil origin, i.e. they are the
descendants of the seed field from the molecular cloud from which the stars were
formed. They are found in $\sim$7\% of all single OBA stars and are mainly
dipolar. Rapid rotation makes it more difficult for fossil fields to reach this
dipolar equilibrium, and this may explain the lack of field detections in
classical Be stars, even though a few examples of rapidly rotating magnetic hot
stars exist.

A few magnetic hot binaries also exist, but magnetism is much less present in
hot binaries than in single hot stars. This might be related to stellar
formation issues: it is more difficult to fragment dense cores when the medium
is magnetic.

These results provide constraints and challenges for formation theories and
simulations to understand the magnetic properties of upper-main sequence stars,
which are very different from those of low-mass and solar-type stars. 

\bibliographystyle{iau307}
\bibliography{Neiner}

\end{document}